\documentclass[preprint]{epl}

\usepackage{psfig}

\title{Efimov states in asymmetric systems}

\author{A.S.~Jensen and D.V.~Fedorov}

\institute{  Department of Physics and Astronomy, University of Aarhus, 
                                            DK-8000 Aarhus C, Denmark }

\pacs{21.45.+v}{Few-body systems}
\pacs{31.15.Ja}{Hyperspherical methods}

\begin{document}

\maketitle

\begin{abstract} 
The conditions for occurrence of the Efimov effect is briefly
described using hyperspherical coordinates. The strength of the
effective hyperradial $\rho^{-2}$ potential appearing for two or
three large scattering lengths is computed and discussed as function
of two independent mass ratios of the three constituent particles.
The effect is by far most pronounced for asymmetric systems with three
very different masses.  One Efimov state may by chance appear in
nuclei. Many states could be present for systems with one electron and
two neutral atoms or molecules. Estimates of the number of states and
their sizes and energies are given.
\end{abstract}

\section{Introduction}

More than 30 years ago Efimov realized that a three-body system could
have a large number of bound states when two or three of the two-body
subsystems simultaneously have (virtual or bound) $s$-states
sufficiently close to zero energy \cite{efi70}. This effect has been
discussed in a number of subsequent publications
\cite{adh82,nie01,jen97,fed94}. External fields can be used to tune
the effective two-body interaction aiming at approaching the zero
energy condition \cite{nie98,kar02,don01}.

The picture describing the Efimov effect is that one particle
effectively has large-distance interaction simultaneously with both
the other two particles building up a coherent wave function fully
exploiting the interactions.  When only one scattering length is large
the effect does not occur since the effective large-distance
interactions only involve two of the particles while the third can
avoid contributing by being far away without interacting.

The atomic helium trimer system is usually considered to be the most
promising candidate for naturally occurring Efimov states
\cite{nie98a,esr96}. It has so far escaped direct experimental
detection \cite{toe02}. Other systems may however be much better
candidates possibly with many Efimov states. The purpose of this
letter is to pin-point both the class of best suited candidates and
the crucial properties optimizing the occurrence conditions.

\section{Basic properties}

The hyperspheric adiabatic expansion combined with the Faddeev
decomposition of the wave function has proven very efficient for
investigations of weakly bound and spatially extended three-body
systems \cite{nie01}. Efimov states can then be computed and without
loss of generality even with the simplifying restriction that the
potentials act only on $s$-waves which are the only contributors at
the asymptotic large distances \cite{jen97}.

The hyperradius $\rho$ defined by 
\begin{equation} \label{e10}
 m \rho^2 \equiv  \frac{1}{ M}  \sum_{i<k} m_i m_k 
({\bf r_i} - {\bf r_k})^2  \;  ,
\end{equation}
where $m_i$ is the mass, $\bf{r}_{i}$ the coordinate of particle $i
\in \{1,2,3\}$ and $m$ is an arbitray normalization mass. Thus $\rho$
is a measure of the average size of the system. After solving the
eigenvalue problem for the remaining (angular) coordinates the method
provides a hyperradial equation with mass $m$ and an effective
potential $U(\rho)$, which for intermediate distances of $\rho$ has
the simple form
\begin{equation}   \label{e20}
 U(\rho) = - \frac{\hbar^2}{2m}\bigg(\frac{\xi^2 + 1/4}{\rho^2} \bigg) \; ,
 \; \; \;   R_e \leq \rho \leq a_{av} \; ,
\end{equation}
where $\xi^2$ is a positive or negative constant depending on the
interactions and the average effective range $R_e$ and average
scattering length $a_{av}$ will be defined later. This potential has
the generic form for the Efimov states.

At distances smaller or comparable with the interaction ranges
$U(\rho)$ is more complicated and generally without divergence as
$\rho^{-2}$. This small distance region provides the scale for the
energies of the possible Efimov states and is otherwise completely
unimportant for spatially extended states. At distances larger than
the scattering lengths the potential has the form \cite{jen97}
\begin{equation} \label{e30}
 U(\rho)  = - \frac{\hbar^2}{2 m \rho^2}  \frac{16}{\pi}
\sum_{i<k} \frac{\sqrt{\mu_{ik}} a_{ik}}{\rho \sqrt{m}}  \equiv 
\frac{\hbar^2}{2 m \rho^2}  \frac{48}{\pi \sqrt{2}} \frac{a_{av}}{\rho }  \;  ,
\end{equation}
where the reduced mass is $\mu_{ik}= m_i m_k/(m_i+m_k)$, the $s$-wave
scattering length is $a_{ik}$ for system $ik$, and the average
scattering length is defined as
\begin{equation} \label{e40}
a_{av} \sqrt{m} \equiv 
\frac{\sqrt{2}}{3} \sum_{i<k} \sqrt{\mu_{ik}} a_{ik} \; . 
\end{equation}
When all three particles are identical the expression for $a_{av}$
reduces to the common value of $a_{ik}$.

The Efimov states are found at intermediate distances as solutions
corresponding to the potential $U(\rho)$ in eq.(\ref{e20}). The wave
function is $K_{i\xi}(\kappa \rho)$, where $K$ is the modified Bessel
function with an imaginary index and the binding energy is $B= \hbar^2
\kappa^2 / (2m)$. By expansion of $K$ for small and large values of
$\rho$ the radial wave function $f_n$ is \cite{nie98a}
\begin{eqnarray}  \label{e50}
 f_n \propto \sqrt{\rho} \sin{\left(\xi \ln (\frac{\rho}
{R_{e}})\right)} \;\; {\rm for} \;\;
 \kappa \rho  < 1 \; , \; \\ f_n \propto \exp(-\kappa \rho) \;\; 
{\rm for} \;\; \kappa \rho > 1   \; ,
\end{eqnarray}
where the zero point for the first oscillation in $\rho$ is assumed to
be $R_{e}$, which in analogy to eq.(\ref{e40}) roughly could be
defined as
\begin{equation} \label{e55}
  R_{e} \sqrt{m} \equiv 
\frac{\sqrt{2}}{3} \sum_{i<k} \sqrt{\mu_{ik}} R_{ik} \; 
\end{equation}
reducing to the common two-body effective radius $R_{ik}$ when all
particles are identical.  The form of the wave function in
eq.(\ref{e50}) is easily found by confirming that $f_n(\rho) \propto
\sqrt{\rho} \rho^{(\pm i\xi )}$ is a solution to the Sch\"{o}dinger
equation with the potential in eq.(\ref{e20}) valid at intermediate
distances where the energy term can be neglected.  The exponential
decrease at large distance occurs for all bound states for distances
$\kappa \rho$ larger than 1, where the effective radial potential
falls off at least as fast as $\rho^{-3}$.

Thus $f_n$ oscillates periodically as $\sin(\xi \ln \rho)$. The number
of Efimov states $N_E$ corresponds then to the number of oscillations
between the average interaction range $R_{e}$ and $48 a_{av}/(\pi
\sqrt{2}) \approx 11 a_{av} $, i.e.
\begin{equation} \label{e60}
 N_E \approx \frac{\xi}{\pi}\ln\left(\frac{11a_{av}} {R_{eff}}\right) \; .
\end{equation}
The energies and sizes of these states are related by
\begin{eqnarray} \label{e70}
  \frac{E_n}{E_{n+1}}
  &=&  \frac{\langle\rho^2\rangle_{n+1}}{\langle\rho^2\rangle_n} =
  e^{2\pi n /\xi}  \; ,
\end{eqnarray}
which reveals the exponential increase of sizes towards infinity and
decrease of energies towards zero, respectively. This behavior
originates from the generic $1/\rho^2$ potential in eq.~(\ref{e20}).

\section{Strength of the potential}

The all decisive parameter $\xi$ must now be determined for systems
where Efimov states may occur. At least two of the three scattering
lengths must be large. The formal analysis in \cite{jen97} leads to
trancendental equations for $\xi$. For identical bosons we obtain the
usual Efimov equation rewritten with real quantities, i.e.
\begin{equation} \label{e80}
 8 \sinh(\xi \pi/6) =  \xi \sqrt{3} \cosh(\xi \pi/2) \; 
\end{equation}
with the solution $\xi \simeq 1.0063$. The result is independent of the
mass of the particles.

For non-identical bosons, still when all three scattering lengths are
large, we find instead
\begin{equation} \label{e90}
 \left(\frac{\xi \cosh(\xi \pi/2)}{2F}\right)^3 - 
  \frac{\xi \cosh(\xi \pi/2)}{2F} \frac{(f^2_1 + f^2_2 + f^2_3)}{F^2} 
 - 2 = 0 \; ,
\end{equation}
where $F = (f_1 f_2 f_3)^{1/3}$ and
\begin{equation} \label{e100}
 f_k =   \frac{\sinh(\xi(\pi/2 - \varphi_k))}{\sin(2\varphi_k)} \; ,
\end{equation}
\begin{equation} \label{e110}
 \varphi_k \equiv \arctan \bigg(\sqrt{\frac{m_k(m_1 + m_2 + m_3)}
 {m_i m_j}}\bigg) \; .
\end{equation}
In general the solution $\xi$ then depends on two ratios of masses,
e.g. $m_2/m_1$ and $m_3/m_1$. When all masses are equal ($\varphi_k =
\pi/3$, $f_i = F = 2 \sinh(\xi \pi /6) /\sqrt{3}$) eq.~(\ref{e90})
reduces to three simpler equations
\begin{equation} \label{e114}
 \frac{\xi \sqrt{3}  \cosh(\xi \pi/2)}{4 \sinh(\xi \pi /6)} =
\left\{ \begin{array}{c}
2 \\ 
\pm 1 
\end{array} \right.  \; , 
\end{equation}
where 2 on the right hand side produce eq.~(\ref{e80}) while $\xi = 0$
is the only real solution for $\pm 1$.

When only the two scattering lengths $a_{jk}$ and $a_{ik}$ (not
$a_{ij}$) are large the Efimov equation for $\xi$ becomes
\begin{equation} \label{e120}
 \xi \cosh(\xi \pi/2) \sin(2\varphi_k) =  2 \sinh (\xi(\pi/2-\varphi_k )) \; ,
\end{equation}
where the solution now only depends on the angle $\varphi_k$ varying
between 0 and $\pi /2$. Still $\varphi_k$ in eq.~(\ref{e110}) in turn
depends on the above two mass ratios.

The conclusion is that the strength of the potential has to be found
by solving eq.~(\ref{e90}) when all scattering lengths are large and
eq.~(\ref{e120}) when only two scattering lengths are large. The
Efimov effect does not occur when only one scattering length is large.
Eq.~(\ref{e80}) is the limit of eq.~(\ref{e90}) when all masses are
equal.  The parameters are the three angles $\varphi_k$, which through
eq.~(\ref{e110}) are functions of two independent mass ratios.

\begin{figure}[!t]
\centerline{\psfig{figure=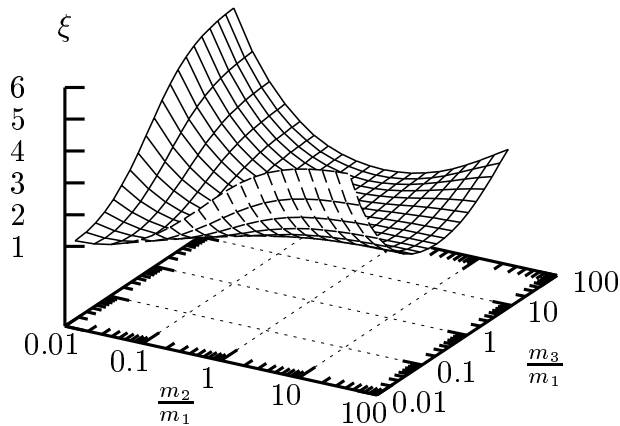,width=8.5cm,%
bbllx=5.2cm,bblly=13.0cm,bburx=12.4cm,bbury=17.9cm}}
\vspace*{0.2cm}
\caption{3d plot of the strength parameter $\xi$, obtained
from eq.~(\ref{e90}) when all three subsystems contribute, as function
of the two mass ratios $m_2/m_1$ and $m_3/m_1$.}
\label{fig1}
\end{figure}

We show in fig.~\ref{fig1} the 3d plot of $\xi$ as function of these
independent parameters.  The smallest values slightly above 1 are
found in a valley passing the symmetric point of three equal masses
and extending towards one very large mass and one moderate mass ratio
around unity. This case with two light masses and one heavy mass
corresponds to the smallest $\xi$ where the Efimov effect is least
pronounced.

On the other hand large values of $\xi$ extending to infinity are
obtained for two heavy and one light mass. This is seen in
fig.~\ref{fig1} when both coordinates are large ($m_1 \ll m_2$, $m_1
\ll m_3$, $m_2 \sim m_3$), and along both axes when the other
coordinate is very small ($m_2 \ll m_1 \ll m_3$ and $m_3 \ll m_1 \ll
m_2$).  Thus moderate values of $\xi$ arise for symmetric systems
while exceedingly large strengths are possible for asymmetric
systems. In all cases the assumption is that all three two-body
subsystems simultaneously have an $s$-state close to the threshold of
binding. This is most likely for a symmetric system of identical
bosons where only one interaction is involved.

\begin{figure}[!t]
\centerline{\psfig{figure=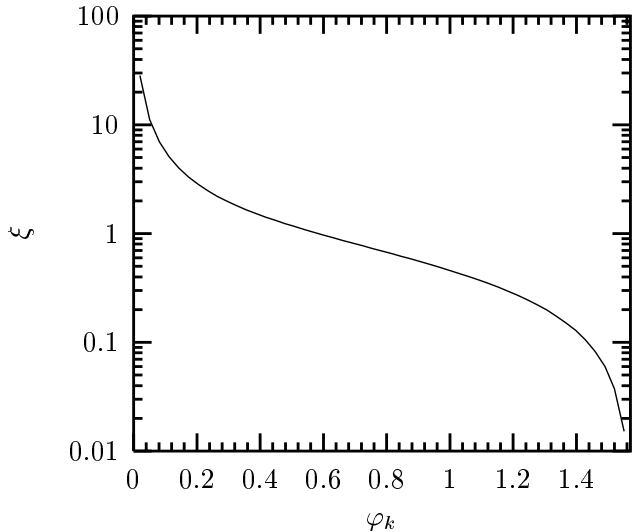,width=8.5cm,%
bbllx=4.7cm,bblly=12.8cm,bburx=12.6cm,bbury=18.8cm}}
\vspace*{0.7cm}
\caption{The strength parameter $\xi$ obtained from eq.~(\ref{e120})
as function of the angle $\varphi_k$.}
\label{fig2}
\end{figure}

For non-indentical particles two tuned subsystems is the least
demanding to exhibit the Efimov effect. If furthermore two of the
clusters are identical particles only one independent requirement of
an $s$-state close to zero is left.  In this case where only two
subsystems contribute (two large scattering lengths) the strength
$\xi$ is found from eq.~(\ref{e120}) as function of the angle
$\varphi_k$ corresponding to the subsystem with the small third
scattering length.  The result displayed in fig.~\ref{fig2} show the
divergence as $1/\varphi_k$ for $\varphi_k \rightarrow 0$ and the
linear convergence to zero as $4(\pi/2 - \varphi_k)/(\pi \sqrt{3})$
for $\varphi_k \rightarrow \pi/2$. When all masses are equal
$\varphi_k = \pi/3$ and eq.~(\ref{e120}) reduces to eq.~(\ref{e80})
with the left hand side divided by two.  The resulting solution $\xi =
0.499$ is roughly two times smaller than 1.0063 obtained when all
three subsystems contribute.

The angle $\varphi_k$ is a function of two mass ratios.  When $m_k$ is
much smaller than both the other masses $\varphi_k$ approaches zero
and $\xi$ is very large. When $m_k$ is much larger than at least one
of the other masses $\varphi_k$ approaches $\pi/2$ and $\xi$ is very
small. Thus extremely large $\xi$ is only possible for two
contributing subsystems when the particle $k$ related to both these
subsystems has a comparatively small mass. The Efimov effect occurs in
all cases but is only pronounced for large $\xi$.

\begin{figure}[!t]
\centerline{\psfig{figure=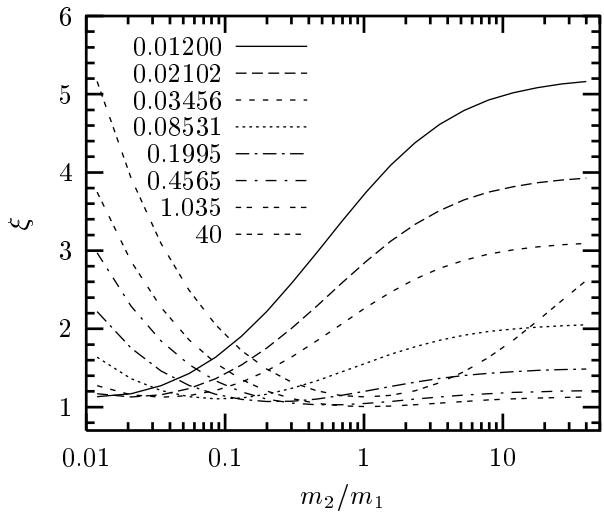,width=8.5cm,%
bbllx=4.7cm,bblly=12.2cm,bburx=13.0cm,bbury=19.0cm}}
\vspace*{0.2cm}
\caption{The strength parameter $\xi$ obtained from eq.~(\ref{e90}) as
function of $m_2/m_1$ for the indicated $m_3/m_1$ values.}
\label{fig3}
\end{figure}

In fig.~\ref{fig3} we show solutions to eq.~(\ref{e90}) corresponding
to three very large scattering lengths for a series of different mass
ratios. The striking features are as described in connection with
fig.~\ref{fig1} that $\xi$ is very small for one heavy and two light
particles, relatively small for similar masses and huge for one light
and two heavy particles. Comparing figs.~\ref{fig2} and \ref{fig3} we
conclude that the mass ratios determine the order of magnitude of the
strength $\xi$ whereas the contribution from the third subsystem is
marginal.

\section{Possible examples}

Large values of $\xi$ favor occurrence of (many) Efimov states, see
eq.~(\ref{e60}). It should be emphasized that at least two scattering
lengths must be sufficiently large to allow the effect in the first
place.  An interaction in one two-body subsystem of longer range than
the generic $\rho^{-2}$ potential prohibits occurrence. For nuclei
this leaves only two neutrons combined with a charged ordinary nucleus
\cite{fed94}. There might be a chance to produce one Efimov state but
the second would not appear as illustrated by $^{11}$Li
($^{9}$Li$+n+n$) where $\xi \approx 0.074$ according to
eq.~(\ref{e70}) corresponds to an increase of the radius by a factor
of $3 \cdot 10^{18}$.

In atomic and molecular physics the possibilities are much better.
For identical particles $\xi$ is mass independent.  The molecular
prototype is the atomic helium trimer $^{4}$He$_3$ with an excited
Efimov state of binding energy 0.18$~\mu$eV and radius about 50~{\AA}.
Correspondingly $\xi \approx 1.0063$ and the radius increase between
neighboring Efimov states is a factor of 22. Equal masses and only two
contributing subsystems change this factor to about 542 implying that
the radius of the second Efimov state then would exceed 1~$\mu$m.  By
far the most favorable case is one light and two heavy identical
particles. One small step in this direction is the asymmetric helium
trimer $^{3}$He$^{4}$He$_2$ with one bound (pronounced halo) state of
binding around 1~$\mu$eV and radius about 13~{\AA} \cite{nie98a,yua98}.

Substituting one helium atom by an alkali atom reduces the two-body
binding energy \cite{kle99} and produce spatially extended three-body
systems like $^{7}$Li$^{4}$He$^3$He and $^{23}$Na$^{3}$He$_2$ with
$\xi \approx 0.255, 0.085$ \cite{yua98}. In $^{6}$Li$^{4}$He$^3$He we
anticipate an Efimov state with a size larger than the 50~{\AA}
expected in $^{4}$He$_3$.  Other combinations with one helium and two
alkali atoms are easily conjectured from the results in \cite{kle99}.
For example $^{3}$He$^{23}$Na$_2$ and $^{3}$He$^{133}$Cs$_2$ give $\xi
\approx 1.22, 2.75$ implying that the second Efimov state only is 13,
3.1 times larger than the first.

Another type of combinations are even more favorable. One charge is
allowed, since the destructive Coulomb-like long-range interaction
still is not present. One electron and two identical atoms (or
molecules) then maximize the $\xi$-value.  It is sufficient with a
large scattering length for the two electron-atom (molecule) systems
and if also the atom-atom (molecule-molecule) contributes the effect
could be even larger.  Then $\xi$ could be as large as 100
(electron-atom (or molecule) mass ratio $\leq 10^{-5}$) and the radius
increase per state from eq.~(\ref{e70}) could be less than 5~\%. The
number of Efimov states within practical reach would increase
dramatically while they still remain relatively stable \cite{pen99}.

However, this scenario requires a large electron-atom scattering
length or equivalently an $s$-state energy very close to zero. Since
the scattering length $a$ at least should be larger than 5-10~{\AA}
the electron binding $\hbar^2/(2m_e a^2)$ should at least be smaller
than 0.1~eV.  This is an established fact for the bound negative ions
Ca, Ti and Sr and furthermore a number of atoms (He, Be, N, Ne, Mg,
Ar, Mn, Zn, Kr, Cd, Xe, Hg, Rn) cannot bind an electron although the
distance from the threshold in general is unknown \cite{and99}.

The three-body system should now consist of the electron plus a pair
of the above atoms chosen as close as possible to the threshold of
$s$-state binding.  Here $^{4}$He$+e+^{4}$He is the obvious
combination but unbound with respect to $^{4}$He$_2$ leaving the
possible Efimov states as excited states with a large decay
probability.  Substituting $^{6}$Li for one of the helium atoms
reduces the atomic binding and $^{4}$He$+e+^{6}$Li is then a promising
candidate.  Other examples like Mg$+e+$Mg, Mg$+e+$Ne, Mg$+e+$Ar,
Mg$+e+$Kr and Mg$+e+$Xe were investigated in \cite{rob99}.  Adding an
electron to the weakly bound two-body systems Ar-Ar, Cd-Cd, Cl-Xe,
Cs-Hg, He-Hg, Hg-Hg \cite{khr99}, also present Efimov candidates, but
pairs like H-H, Ca-Ca, Li-Li, Na-Na could be more suitable.

Instead of atoms we could try to use molecules with very small
electron binding, combining them pairwise and adding an electron, see
\cite{khr99,rie02} for candidates. If this should create Efimov states
the internal structure of the molecules is not allowed to change
within the pair. This might be possible but requires a separate
investigation.

The Efimov conditions of two zero energy subsystems can be fulfilled
without lower lying bound two-body states, but it is much more
probable to encounter systems where the state of zero energy is an
excited two-body state.  The resulting Efimov states are clearly more
unstable and their structure more difficult to study directly.
However, signals should show up in scattering experiments
\cite{hel96,nie02}. The properties of the Efimov states are in any
case determined by the scattering lengths and the size of $\xi$ as
discussed above.

\section{Conclusion}

The strength of the attraction in the effective hyperradial
$\rho^{-2}$ potential for three particles is for large scattering
lengths determined by simple equations depending only on two
independent mass ratios.  We survey this mass dependence and provide
information about favorable mass combinations. The larger the strength
the more pronounced is the Efimov effect.  The density of Efimov
states increases with the square root of the strength which for
existing mass combinations can vary by several orders of magnitude.
Two contributing subsystems for one light and two heavy particles is
much more favorable than three contributing subsystems with similar
masses.  Careful choices of the three particles can then
simultaneously optimize the Efimov conditions of large strength and
large scattering lengths. This may allow detection of several or many
members in a series of Efimov states or alternatively allow detection
of the first Efimov state in cases when the conditions are less well
fulfilled.  We suggest a number of possible combinations of an
electron and two neutral atoms or molecules, or alternatively one
light atom and two heavy atoms or molecules.

\end{document}